# CABS-flex 2.0: a web server for fast simulations of flexibility of protein structures

Aleksander Kuriata[1,&], Aleksandra Maria Gierut[1, 2, &], Tymoteusz Oleniecki[1,3], Maciej Paweł Ciemny[1,4], Andrzej Kolinski[1], Mateusz Kurcinski[1]*, Sebastian Kmiecik[1]*

[&]equal contribution
*emails of corresponding authors: sekmi@chem.uw.edu.pl, mkurc@chem.uw.edu.pl

[1]Biological and Chemical Research Centre, Faculty of Chemistry, University of Warsaw, Warsaw, Poland
[2]Faculty of Physics, Astronomy and Applied Computer Science, Jagiellonian University, Krakow, Poland
[3]College of Inter-Faculty Individual Studies in Mathematics and Natural Sciences, University of Warsaw, Warsaw, Poland
[4]Faculty of Physics, University of Warsaw, Warsaw, Poland

**ABSTRACT**
Classical simulations of protein flexibility remain computationally expensive, especially for large proteins. A few years ago, we developed a fast method for predicting protein structure fluctuations that uses a single protein model as the input. The method has been made available as the CABS-flex web server and applied in numerous studies of protein structure-function relationships. Here, we present a major update of the CABS-flex web server to version 2.0. The new features include: extension of the method to significantly larger and multimeric proteins, customizable distance restraints and simulation parameters, contact maps and a new, enhanced web server interface. CABS-flex 2.0 is freely available at http://biocomp.chem.uw.edu.pl/CABSflex2

**INTRODUCTION**

Dynamics of protein structures defines their biological functions. Because the experimental investigation of protein flexibility is often difficult or impossible, computational approaches play a significant role in this field. Simulations of biologically relevant protein fluctuations remain computationally demanding (using classical modeling tools of atomistic resolution) and often require supercomputer power. An inexpensive alternative is using coarse-grained simulation models combined with the reconstruction of predicted structures to all-atom representation (1). In 2013, we developed the CABS-flex web server for fast simulations of near-native dynamics of globular proteins (2). The CABS-flex method was shown to be a computationally efficient alternative to the classical, all-atom molecular dynamics (3). We also demonstrated that fluctuations of protein residues obtained from CABS-flex are well correlated to those of NMR ensembles (4). The CABS-flex method is also used as a part of the AGGRESCAN3D method for the prediction of protein aggregation propensities (5) (AGGRESCAN3D employs CABS-flex to include the influence of dynamic protein structure fluctuations on aggregation propensity). Moreover, the CABS-flex methodology is a component of the CABS-dock method for protein-peptide docking (6-8), which enables significant flexibility of a peptide and a protein receptor during explicit simulation of peptide binding.

The CABS-flex provides an alternative to other efficient methods of generating protein residue fluctuation profiles, such as sequence-based predictors of protein disordered regions or other coarse-grained approaches. In comparison to sequence-based predictors (9), CABS-flex is better adapted to detecting non-obvious dynamic fluctuations, for example within the well-defined secondary structural elements that could be biologically relevant.

Other computational tools use normal mode analysis (NMA) or various kinds of coarse-grained models (10-14) Their limitations depend on their particular design, especially on the simplifications assumed, and on the modeled system (for example NMA is well suited only for certain systems (15)).

In this work, we present a major update of the original CABS-flex, which significantly extends its capabilities. The CABS-flex 2.0 has been equipped with three major feature upgrades:
- the limitations of protein structure input have been extended from 400 to 2'000 amino acids, and from single-chain proteins only to proteins consisting of up to 10 chains;
- the panel of customizable simulation parameters and options that enable a deeper control of the simulation process, including user-defined modifications of distance restraints;
- protein contact maps for simulation trajectory (presenting frequency of residue-residue contacts during simulation) and for 10 representative protein models.

According to the users' feedback, the major drawbacks of the original CABS-flex server were the protein size limitations (restricted to only single chain proteins shorter than 400 amino acids). The significant extension of protein size limits (described above) was possible thanks to dedicating significantly larger computational resources to the web server jobs and rewriting the CABS-flex code.

The original CABS-flex server did not allow for modifications of the simulation settings, while the CABS-flex 2.0 features customizable simulation parameters and options that enable users to tailor the simulation according to their requirements. The customizable simulation parameters include (among others): temperature, simulation length and distance restraints. The introduced versatility makes the CABS-flex server suitable for performing complex simulations of: proteins with disordered regions of significant length, flexible loops or simulations including user-defined distance restraints.

Additionally, the newly designed interface of the CABS-flex 2.0 server provides new and intuitive input forms, as well as extended output panels for interactive result analysis. A new feature of contact maps facilitates the results analysis and provides direct insight into intra-molecular interactions of a modeled protein.

**MATERIALS AND METHODS**

*CABS-flex modeling protocol*
The original CABS-flex server was described in details elsewhere (2). CABS-flex modeling results were validated against all-atom MD data (3) and NMR experimental structures (4). The method was also extensively validated as a component of modeling tools for predictions of protein solubility (5) and flexible peptide docking (6-8). The key principles of CABS-flex 2.0 remain similar to these of the original web server. The overview of the method pipeline is presented in the Figure 1.

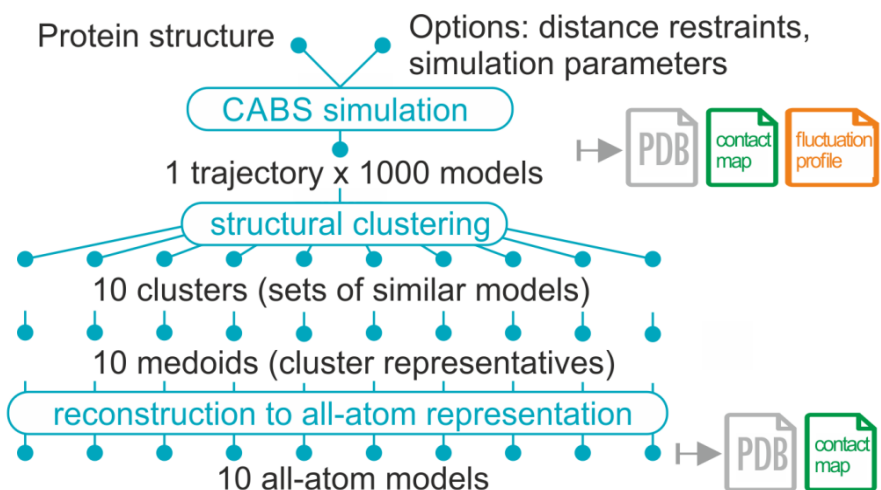

**Figure 1.** CABS-flex 2.0 pipeline. In the first modeling step, the CABS-flex uses CABS coarse-grained protein model as an efficient simulation engine (the CABS model design and applications have been recently reviewed (1)). Next, a protein dynamics trajectory is clustered into 10 representative models, which are automatically reconstructed to all-atom representation.

**DESCRIPTION OF THE WEB SERVER**

**Input data**

The only required input is the protein structure. It may be either provided as a PDB code or as an uploaded PDB format file. In both of these cases, all the residues must comprise a complete set of backbone atoms (i.e. N, Cα, C and O). The input protein structure may contain multiple chains. However, (due to limited computational resources) each chain must be no longer than 2000 residues. Additionally, the CABS-flex 2.0 allows providing optional input information's:

- Chain name(s) provided as a multi-letter code used to select specified chains from the uploaded PDB file; for example, 'ABCD' allows to select A, B, C and D chains from the PDB file;
- Project name, which will appear in the queue list, unless the option "Do not show my job on the results page" is used. If not provided, the name will be replaced with random hashcode;
- E-mail address, which will be used by the server to send an email notification about job completion

**Advanced input options**

In default mode, CABS-flex uses a set of distance restraints and simulation settings defined in work of Jamroz et al. (3). The default settings and restraints were optimized to provide the best possible convergence between CABS-flex simulations and a consensus picture of protein fluctuations in aqueous solution derived by all-atom Molecular Dynamics (MD) simulations (of 10 nanosecond length, with different force fields) of globular proteins. CABS-flex predictions of protein fluctuations were also shown to be well correlated to fluctuations seen in NMR ensemble (4).

The advanced options enable modification of the default settings according to user's needs and information about the modeled system. The advanced options are organized under three dropdown panels: protein structure input options, additional distance restraints, advanced simulation options; which are described below.

*Protein structure input options*

This panel allows user to customize Cα– Cα restraints imposed on the protein residues. By default, the CABS-flex generates a set of distance restraints in order to keep the modeled protein near input (near-native) conformation. Restraints are generated according to the scheme defined by the 'Mode', 'Gap', 'Minimum' and 'Maximum' fields. 'Protein flexibility' further modifies the generated list of restraints.

- 'Gap' sets the minimum distance along the protein chain for two residues to be bound with a restraint (default: 3), i.e. Gap = 3 means that residue number 15 cannot be restrained with residues from number 12 to 18
- 'Minimum' and 'Maximum' fields define the minimal and maximal length of the restraint in Angstroms. In other words, restraints will be automatically generated only for residues within these distances (default values are 3.8 and 8.0, for minimum and maximum, respectively)
- 'Mode' tells the algorithm to keep only these restraints for which at least one (SS1) or both (SS2) restrained residues are assigned a regular secondary structure (helix or sheet). 'Mode' = 'all' generates restraints for all residues and 'none' generates no restraints

The option 'Protein flexibility' can be used to assign to each of the residues a single value of flexibility. Flexibility is a number between 0 (fully flexible) and 1 (rigid), which is used to assign weights to distance restraints. The restraint weight is equal to the lower of two flexibilities of restrained residues - i.e. restraint between residue with flexibility = 0.75 and residue with flexibility = 0.23 will have weight = 0.23. 'Protein flexibility' can be set in four ways:

- single number in the 'Protein flexibility field' sets uniform flexibility for all residues equal that number (default 1.0)
- 'bf' keyword in the 'Protein flexibility' sets the flexibility equal to the beta factor column in the input PDB file. User can prepare custom input file with desired flexibility values assigned to the residues and use it on input
- 'bfi' keyword in the 'Protein flexibility' field allows to use real beta factors from the PDB input file. Crystallographic beta-factors have somewhat opposite meaning to the CABS flexibility, but can be interpreted to serve as such
- flexibility can be also set by uploading a custom file with rules assigning flexibility to single residues or ranges of residues at once

The option 'Manually edit protein restraints' provides an additional web-based tool for manual restraints edition. When selected 'Yes', the option redirects the user to a new web page before the job is submitted. The CABSflex 2.0 generates a set of restraints to be used in simulation based on selected options. Each of the generated restraints can be manually edited (including its strength and distance) or deleted. Additionally new restraints between any of the protein's residues can be created and further modified.

*Additional distance restraints*

This form allows to insert additional (to those automatically generated) restraints imposed on modeled protein structure. Top panel is for Cα– Cα restraints and the bottom panel is for side chain - side chain restraints (SC-SC). Restraints can be added either through the text box or from uploaded text file. Each line represents one restraint and should have the following syntax:

residue1_id residue2_id distance weight, where residue_id has a form of Number:Chain as in 123:A

*Advanced simulation options*

This panel allows for modification of the parameters controlling the simulation. 'Number of cycles'($N_{cycle}$) field sets the total number of models saved in the trajectory to be equal to 20 x $N_{cycle}$ - i.e. setting $N_{cycle}$ = 50 results in 20 x 50 = 1'000 models in the trajectory. It is worth noting that not all of the models generated during the CABS-flex simulation are written to the trajectory. The next field - 'Cycles between trajectory frames' ($N_{skipped}$) - sets the number of models skipped upon saving models, in other words for $N_{skipped}$ = 100 every hundredth model will be saved. This field also indirectly sets the total number of models generated - i.e. for $N_{cycle}$ = 50 and $N_{skipped}$ = 100, the total number of generated models equals to 20 x 50 x 100 = 100'000, 1'000 of which will be written to the trajectory. For more details refer to the wiki pages of the standalone package of the CABS-flex, available at https://bitbucket.org/lcbio/cabsflex

The dimensionless reduced temperature in the CABS model cannot be straightforwardly linked to the real temperature. Its role, however, is similar - it serves as a parameter controlling the total energy of the modeled system. The higher the temperature, the more mobile are the atoms, which results in larger fluctuations. In the CABS model, T = 1.0 is usually close to the temperature of the crystal (native state), T = 2.0 typically allows for complete unfolding of unrestrained small protein chains.

**Figure 2. Input panels.** By default, all panels except the 'Input data' are hidden and can be displayed by clicking on the selected tab.

**Output files and data**

Once the job is completed three additional tabs are added to its web page: "Models", "Contact maps" and "Fluctuation plot".

*Models tab*

This section is used to display three-dimensional structures of 10 final models (see Figure 3A) and to provide the structure data. An interactive molecular viewer enables visual analysis of the obtained results. "model_all" set consists of all 10 final models, therefore its visualization shows the structure heterogeneity present in the final models. Additional viewing options are available, such as toggling between surface and cartoon representation and rotating the protein molecule about a vertical axis. User may change what is being displayed in the viewer by selecting from the side menu one of the ten final models or the trajectory – all final models in superposition. Each final model as well as the trajectory may be downloaded as PDB files.

*Contact maps tab*

The "Contact maps" tab provides a detailed view on the protein's residue-residue interaction pattern (see Figure 3B). Central panel displays an interactive contact map between residues. Each dot in the map represents an interaction within a pair of residues. Its color depends on the frequency of occurrence of this particular interaction in the models set that was used to generate data for the map. User may select in the side tab, which set of structures is used to populate the map, by clicking on the 'View' button, next to one of the twelve preset sets. Sets named model_1-10 are singletons containing just one of the final models, therefore be selecting one of them a binary (contact frequency is either 0 or 1) contact pattern for this model is displayed. "Trajectory" set contains all models saved in the trajectory during simulation (by default 1000 models) - the "trajectory" contact map displays which contacts were most frequent throughout the simulation. "model_all" set consists of all ten final models, therefore its map shows the most conserved structural features present in the final models. Maps are available for download as a graphic (svg format) and a text files (both in zip files under Download buttons). Hovering on any of the square displays a rectangle in the upper-left corner with more detailed information about the selected contact. The color range of the contact map is adjustable in the 'Options' panel. Any transformation of the contact map (translation, scaling) can be reversed by clicking on the 'Reset' button below it.

### *Fluctuation plot tab*

"Fluctuation plot" tab provides an interactive 2D plot presenting residue-wise fluctuations recorded throughout the simulation. Fluctuations are calculated as the RMSF after global superposition. Both graphics (svg) and numerical data (csv) are available for download. For multichain proteins a separate plot is generated for each chain. They can be displayed by selecting it in the 'Chains' panel.

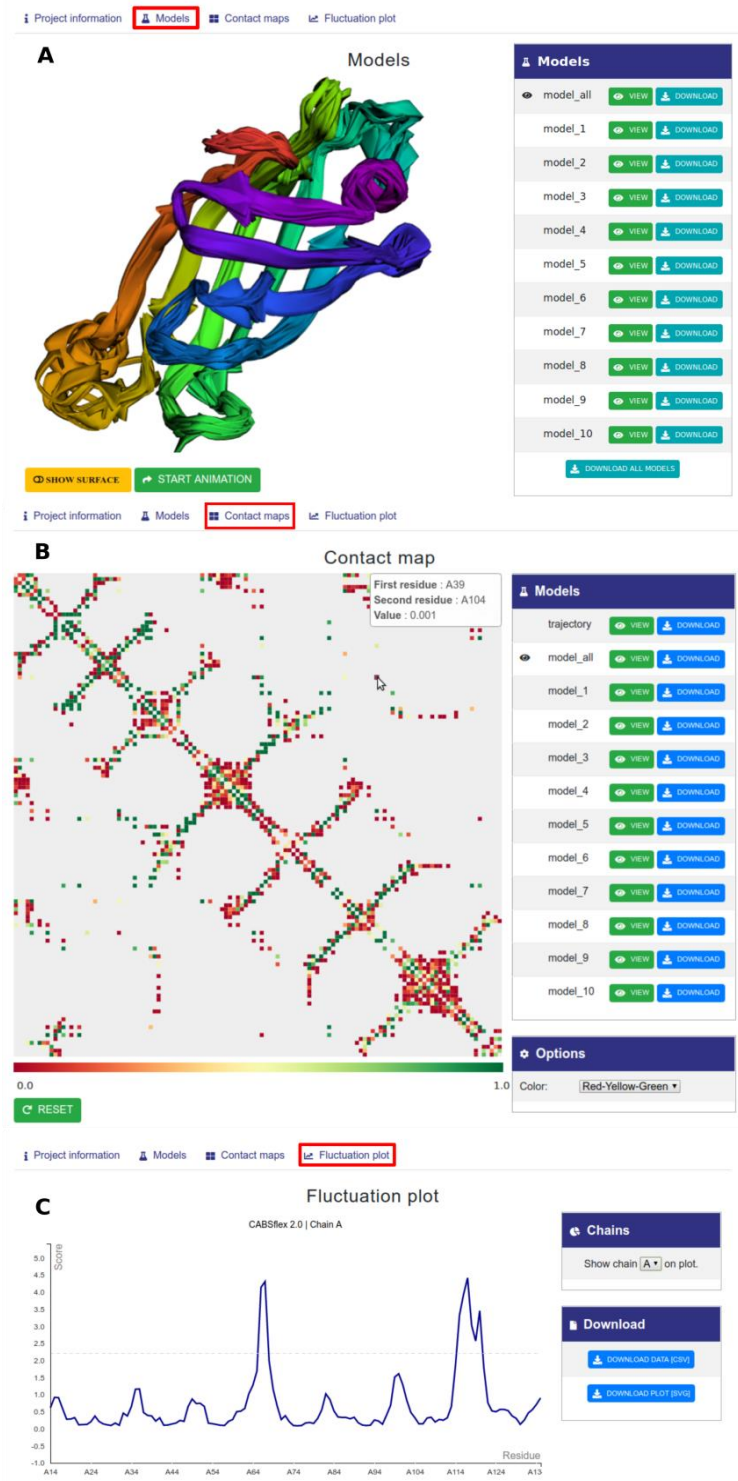

**Figure 3. Example view of output tabs for streptavidin protein (with default CABS-flex settings, 2RTM PDB was used as the input).** (A) 'Models' tab. Left panel shows three-dimensional visualization of 10 models. Right panel provide buttons to display and download the selected model or all 10 models (model_all). Below: buttons to modify the surface and rotate the model. (B) 'Contact maps' tab. The tab shows: matrix with contacts with color scale below (notice that pointing the cursor at the contact between residue 39 and 104 in chain A invoked an information balloon with the contact information). Right panels provide buttons to display and download maps (in a graphic svg format and a text files, both in zip files under Download buttons), and Options section that allow changing the color range of contacts. (C)

'Fluctuation plot' tab. Plot shows residue fluctuation profile (RMSF) for selected protein chain. Right panels allow selecting a protein chain and downloading the displayed data.

## SERVER ARCHITECTURE AND DOCUMENTATION

The CABS-flex 2.0 server is equipped with a HTML web interface dynamically generated with Flask framework and jinja2 templating engine. Validated user-provided data are added to the MySQL database. The job is either started, if there are free server resources available, or put in the queue to wait for execution. The server notifies the user on the progress of the computation, reporting the job status ('pending', 'in queue', 'running' and 'done'). The molecular visualization is executed using 3Dmol library (HTML5/Javascript). The interactive contact maps and plots are created with D3.js library (HTML5/Javascript). Information about proteins is downloaded from the PDB using RESTful services. The CABS-flex 2.0 website runs on the Apache2 server and MySQL database for user queue storage. As the simulation engine, the server uses newly developed CABS-flex standalone package, which repository is available at https://bitbucket.org/lcbio/cabsflex. CABS-flex 2.0 server is free, open to all users and there is no login requirement.

The documentation of the CABS-flex 2.0 is available online under 'How to' subpage.

## SUMMARY

In this work, we developed easy-to-use CABS-flex 2.0 web server interface for efficient simulations of large-scale (large in context of protein size and timescale of observed fluctuations) structure fluctuations of proteins and protein complexes. CABS-flex 2.0 is based on the coarse-grained simulations of protein motion that has been successfully used in CABS-flex 1.0 server (2,4) and other modeling tools (5-8). The present server allows simulations of much larger systems and more elaborate (when required) control of flexibility of amino acid chains than the previous implementation. In particular, one can indicate more restricted regions of modeled proteins as well as those that may adopt very different structures than the one used in the input. Produced mobility profiles and fluctuation trajectories can be used in studies of biomolecular processes. In particular, one may use CABS-flex profiles to: identify the most mobile structural fragments, generate structures of protein receptors for molecular docking (as the alternative to available experimental structures), study allosteric equilibria and many other tasks that require large-scale dynamics data.

## FUNDING
The authors acknowledge support from the National Science Center (NCN, Poland) Grant [MAESTRO2014/14/A/ST6/00088]